\documentclass[%
reprint,
amsmath,amssymb,
aps,
]{revtex4-2}

\usepackage{graphicx}
\usepackage{dcolumn}
\usepackage{bm}

\begin{document}
	\renewcommand{\arraystretch}{1.5}
	
	\title{Coalescing Clusters Unveil New Regimes of Frictional Fluid Mechanics}
	
	\author{Haicen Yue}
 	\altaffiliation[Now at ]{Department of Physics, the University of Vermont}
 	\email{haicen.yue@uvm.edu}
        \author{Justin C. Burton}%
	\author{Daniel M. Sussman}
	\email{daniel.m.sussman@emory.edu}
	\affiliation{%
		Department of Physics, Emory University, Atlanta, GA 30322
	}%

	\date{\today}
	
	\begin{abstract}
		The coalescence of droplets is essential in a host of biological and industrial processes, where it can be both a crucial limiting process and a convenient tool to probe material properties. Controlling and understanding this phenomenon relies on theoretical models of droplet coalescence, but classical solutions for the time evolution of contacting liquids are typically based on tractable limiting physics, such as analytical solutions to the Stokes equation. Discrepancies between these classical solutions and experimental results abound. By combining computational and theoretical analyses, we show that there is an unexplored family of ``dry hydrodynamic'' or ``frictional'' coalescence processes: those governed by a highly dissipative coupling to the environment. This leads to a novel universality class of coalescing behavior, with new scaling laws and time-invariant parameterizations for the time evolution of coalescing drops. To demonstrate this, we combine particle-based simulations and both continuum and boundary-integral solutions to hydrodynamic equations, which we then understand in the context of a generalized Navier-Stokes-like equation. Our work suggests a new theoretical basis for further studies of coalescence, as well as other fluid-like phenomena in these friction-dominated systems, and significantly alters the interpretation of related experimental measurements.
	\end{abstract}
	
	\maketitle
	
	
	\section{Introduction}
	Droplet coalescence driven by surface tension is a visually striking process, and one that is crucial in applications as diverse as  meteorology \cite{grabowski2013growth}, the food \cite{tan2021application} and pharmaceutical industries \cite{barkat2011basics},  inkjet printing \cite{sun2015recent}, and the treatment of wastewater \cite{hussein2019application} and oil spills \cite{zhu2022recent}. In some systems controlling the coalescence process itself is the goal, and in others coalescing behavior can be used to infer material properties. For instance, in recent years the theory of coalescence has served as the basis for studying fusion in liquid-like biological systems such as aggregates of living cells  \cite{grosserCellNucleusShape2021a,koshelevaCellSpheroidFusion2020a, oriola2022arrested}, shedding light on critical biological processes in embryonic development, cancer invasion, and tissue engineering. In the case of simple liquids at low Reynolds number, droplet coalescence is well-understood, as conformal mapping techniques can be used to find analytical solutions to the relevant Stokes equation \cite{hopperCoalescenceTwoViscous1993,hopperCoalescenceTwoViscous1993a}. 
	
	A general understanding of the coalescence of more complex liquid or liquid-like systems is made difficult by various system-specific features, e.g., viscous (and viscoelastic) effects, entanglements in polymers, or the active motion of living cells. However, many of these more complex systems share a common characteristic: their constituents are strongly coupled to a highly dissipative background, resulting in a lack of momentum conservation. Examples include cells moving on a rigid two-dimensional substrate or embedded in the Extracellular Matrix (ECM), fluid flows through porous media, and flows between two closely spaced walls. A common framework classifies hydrodynamic systems as ``dry'' (when momentum is not conserved) or ``wet'' (when it is), and it is well known that these limits can result in qualitatively different collective behaviors \cite{marchetti2013hydrodynamics,doostmohammadi2016stabilization}. The boundary between wet and dry is often not clear-cut: the same system can be regarded as dry or wet depending on whether hydrodynamic effects can be neglected, or whether momentum can be regarded as a fast variable on the length and time scales of interest. For example, although bacteria suspensions are usually regarded as wet, the turbulent flows in dense bacteria suspensions are well explained by dry models with overdamped dynamics \cite{wensink2012meso}. This distinction is also important in non-living systems, such as colloidal suspensions of particles much denser than the solvent: there is a clear time scale separation between the dynamics of the colloids and the solvent, and (although the terminology is not typically used) the colloids can be approximated as dry \cite{hauge1973fluctuating,snook2006langevin}. In these dry systems, the movement of the constituents is often modeled via a Langevin equation, in which the degrees of freedom of the background are coarse-grained into fluctuation and dissipation terms \cite{espanol2004statistical}. 
	
	\begin{figure*}[ht]
		\centering
		\includegraphics[width=\textwidth]{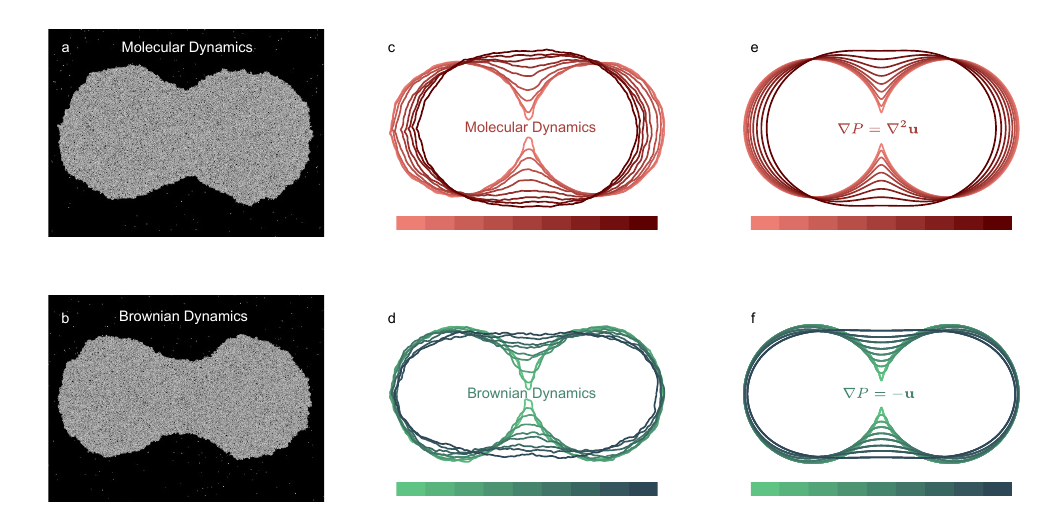}
		\caption{\textbf{Droplet coalescence as a function of microscopic dynamics.} Representative snapshots of our discrete simulations evolving according to (a) Molecular or (b) Brownian dynamics. (c,d) Contours in the corresponding simulations at different times during the coalescence process, ensemble-averaged over 5 (MD) and 3 (BD) simulations. (e,f) Contours at different times for solutions to the Stokes equation (e) and Darcy's Law (f) for coalescing droplets. In (c-f) darker colors are used to represent later time points.}
		\label{fig:Contours}
	\end{figure*}
	
	Although coalescence is just one of many hydrodynamic phenomena, we believe that this process is an ideal testbed for studying hydrodynamic differences between wet and dry systems: it is experimentally well-studied across multiple hydrodynamic regimes, and it is used as a representative process of intense interest in living systems such as bacteria colonies \cite{ponisch2018pili,welker2018molecular} and collections of cells \cite{grosserCellNucleusShape2021a, koshelevaCellSpheroidFusion2020a, oriola2022arrested}. Despite this ubiquity, there has been no systematic investigation of how coalescence is changed in even the simplest of dry systems. The fundamental question of whether an effective change in microscopic dynamics -- e.g., from Newton's law to a Langevin description -- has a non-trivial effect on coalescence dynamics remains unclear. As a result, many studies on potentially dry systems still rely on low Reynolds number hydrodynamic theories for simple-liquid coalescence as a basis for comparing both experiments and agent-based simulations \cite{flennerKineticMonteCarlo2012,brangwynne2011active, caragineSurfaceFluctuationsCoalescence2018a,grosserCellNucleusShape2021a,koshelevaCellSpheroidFusion2020a}. These theories are then used to infer physical properties, such as the effective surface tension or viscosity, in otherwise difficult-to-measure systems. Examples of this include measuring characteristic coalescence times in cellular systems \cite{brangwynne2011active,grosserCellNucleusShape2021a} or observing both coalescence dynamics and surface tension fluctuations in the fusion of nucleoli \cite{caragineSurfaceFluctuationsCoalescence2018a}. Deviations from the theoretically anticipated behavior are then used to infer additional complexities, such as the presence of heterogeneity in cellular spheroids \cite{grosserCellNucleusShape2021a,koshelevaCellSpheroidFusion2020a}.

	In this work we demonstrate that the difference in microscopic dynamics -- i.e., the difference between wet and dry systems -- leads to coalescence behaviour in a different universality class, as schematically illustrated in Fig.~\ref{fig:Contours}. This first suggests that specific experimental data from many disparate dry systems should be reinterpreted in the context of frictional rather than Stokesian coalescence, and further highlights the importance of considering frictional fluid mechanics as a distinct starting point from a more traditional approach in any system with strong coupling to a dissipative, momentum-non-conserving background.

	The rest of this manuscript is organized as followed. We begin by describing particle-based simulations in which Lennard-Jones particles in two coalescing droplets evolve in time according to either standard Molecular Dynamics or Brownian Dynamics. Details of the equations of motion used in this work and their applicability and limitations are in Sec.~\ref{sec:Methods}. In this section we also approach the problem at a hydrodynamic level, adding an additional frictional term to a Navier-Stokes-like description \cite{orts2020dynamics}. The highly frictional extreme of this continuum model is intimately linked to Darcy's Law, a mathematical model of fluid flow through porous media, but here in the context of a compact system governed by surface tension. By solving Darcy's law with Boundary Integral Methods in the context of coalescence, we convincingly show that this equation, instead of Stokes' equation for viscous simple  liquids, should be used to describe the coalescence of highly frictional fluids. In both particulate simulations and hydrodynamic solutions we observe fundamental differences in the overall qualitative shape evolution of the merging droplets (Sec.~\ref{sec:shape}). In Sec.~\ref{sec:scale} we report unexpected scaling laws characterizing the growth of the neck in time near the critical (initial contact) point, suggesting that coalescence in highly frictional environments is governed by a novel universality class. In Secs.~\ref{sec:shape} and \ref{sec:scale} we additionally analyze the extended hydrodynamic model presented in Sec.~\ref{sec:Methods} by systematically comparing the effect of inertial, viscous, and frictional terms and provide a way to qualitatively predict the coalescing behavior of frictional systems with different parameter combinations including system size, density, viscosity, frictional coefficient, and surface tension. Finally, in Sec.~\ref{sec:future}, we discuss potential extensions of this work and in Sec.\ref{sec:significance}, we discuss the influence this work can have on advancing the understanding and shaping future research directions in diverse particle and biological systems via several illustrative examples.

	\section{Methods}
	\label{sec:Methods}
	\subsection{Particle-based Simulations}
	In this work, we simultaneously study coalescence under three different microscopic evolution equations. The first is  standard Molecular Dynamics (MD), in which we numerically integrate Newton's laws,
	\begin{equation}
		m_i\frac{d\textbf{v}_i}{dt} = \sum_j\textbf{F}_{ij}.
	\end{equation}
 
	The second is Langevin Dynamics (LD), in which particles evolve according to 
	\begin{align} \label{eq:LD}
		&m_i \frac{d\textbf{v}_i}{dt} = \sum_j\textbf{F}_{ij}-\lambda\textbf{v}_i +\textbf{F}_R\\
		&\langle\textbf{F}_R\rangle=0, \quad \langle \lvert\textbf{F}_R\rvert^2 \rangle = 4k_BT\lambda/\delta t,\nonumber
	\end{align}
	where $\lambda$ is a frictional term and the $\textbf{F}_R$ are time-uncorrelated Gaussian noise terms. We note that in many-body colloidal systems, a more general form of the Langevin Equation has a friction term as $-\sum_j \bm{\lambda}_{ij} \textbf{v}_j $, in which the hydrodynamic interactions are embedded in the friction tensor $\bm{\lambda}_{ij}$. Only when the colloidal suspension is dilute enough can the friction tensor be strictly reduced to a diagonal form, $\lambda\bm{I}$. Unfortunately, the calculation of the friction tensor in dense colloidal systems is quite challenging: the real interactions are both long-ranged and not pair-wise additive. 
 
 The third set of microscopic evolution equations we use are Brownian Dynamics (BD), a limiting version of Langevin dynamics with
	\begin{align}
		&\lambda\frac{d\textbf{r}_i}{dt} = \sum_j\textbf{F}_{ij}+\textbf{F}_R\\
		&\langle\textbf{F}_R\rangle=0, \quad \langle \lvert\textbf{F}_R\rvert^2 \rangle = 4k_BT\lambda/\delta t.\nonumber
	\end{align}
	
	These three evolution equations represent the standard molecular dynamics of classical microscopic degrees of freedom (MD) and two extensions of it meant to represent a coupling between degrees of freedom and a dissipative background (LD and BD). Langevin and Brownian dynamics are most commonly used to approximate the effect of mesoscopically large objects (e.g., colloids) interacting with a viscous solvent -- in this setting they are most appropriate when applied to systems in which solvent-mediated hydrodynamic interactions between the degrees of freedom can be neglected. For instance, when the pair-wise interactions between the colloidal particles are very strong, the many-body hydrodynamic effect can be relatively small after averaging over all the neighbors. As a result, despite lacking a fully rigorous basis, the equations for LD and BD above have been widely used in simulations of concentrated colloidal systems on large enough scales of observation so that the movement and collisions of the small solvent molecules can be coarse-grained to a mean-field frictional effect, and have provided useful insights into various colloidal phenomena, including glass formation and glassy dynamics \cite{lowen1991brownian,flenner2015fundamental}, rheology \cite{foss2000brownian,treffenstadt2020memory}, and flows through porous media \cite{chavez2008diffusion}. 
 
 These dissipative dynamics are also important when considering degrees of freedom directly coupled to a frictional background that can serve as a momentum sink, such as cells crawling on a substrate or interacting with a fibrous background network. In these systems the momentum of cells is simply not a conserved quantity at any scale compariable to or above the individual cell scale. As such, Brownian and Langevin dynamics have been used extensively in the simulation of the collective movement of cells, although additional active forces and more complicated cell-cell interactions are often also implemented \cite{camley2017physical}.
	
Despite all the aforementioned potential complexities, in this work we will focus solely on the influence of damping on coalescence dynamics, and demonstrate that this change \emph{alone} drastically alters the geometry and dynamics. We choose the above equations and a 2D binary Lennard-Jones system $A_{65}B_{35}$ with which we can simulate large enough systems. Previous studies on simple liquid coalescence have shown that the dynamics of neck growth in 3D is asymptotically equivalent to the 2D case \cite{eggersCoalescenceLiquidDrops1999a}, and simulations with different inter-particle potentials \cite{pothier2012molecular,heinenDropletCoalescenceMolecular2022a} have all produced outcomes consistent with theoretical predictions. These suggest that the 2D Lennard-Jones system is a reasonable starting point. Specifically, $U(r) = 4\epsilon\big[\big(\frac{\sigma}{r}\big)^{12}-\big(\frac{\sigma}{r}\big)^6\big]$, where $\sigma_{AA} = 1.0$, $\epsilon_{AA} = 1.0$,$\sigma_{AB} = 0.8$, $\epsilon_{AB} = 1.5$,$\sigma_{BB} = 0.88$, $\epsilon_{BB} = 0.5$ and $r_{cut}=2.5$, as in Ref.~\cite{bruningGlassTransitionsOne2008}. The simulations are run at temperature $kT=0.35$, which is in the liquid phase, with default values of $m=1$ and $\lambda=1$ and a time step size of $\delta t = 0.0002$ in natural Lennard-Jones units.  The MD simulations are done in the canonical (NVT) ensemble with a Nos\'{e}-Hoover thermostat. All of these particle-based  simulations are run with the \textit{HOOMD-blue} \cite{andersonHOOMDbluePythonPackage2020} simulation package on XSEDE \cite{towns2014xsede}. 
	
	\subsection{Continuum Equations}
	The continuum equations we use for the simple viscous liquid case and frictional liquid case (corresponding to the MD simulation and BD simulation results respectively) are Stokes' equation and Darcy's law, supplemented with the correct surface-tension-mediated boundary conditions and assuming an incompressible fluid:
	\begin{align}
		&\nabla P = \mu\nabla^2\textbf{u} \textrm{ (viscous-dominated), or} \\
		&\nabla P = -\zeta\textbf{u} \textrm{ (friction-dominated), with} \\
		&P_{ext}-P_{int} = -\gamma\kappa. 
	\end{align}
	Here $\mu$ is the viscous coefficient, $\zeta$ is the frictional coefficient, $\gamma$ is the surface tension and $\kappa$ is the boundary curvature. The choice of Darcy's law is based on the idea that in both BD and fluid flow through pores, the momentum of the flow is quickly dissipated to its surroundings and indeed, we will later show that Darcy’s Law is an extreme, overdamped limit of a more general Navier-Stokes-like description.
	
	Denoting the initial radius of one droplet by $R$, we rescale length, mass, and time via $\{L=R$, $T = R\mu\gamma^{-1}$, $M = \gamma T^2\}$ and $\{L=R$, $T = R^3\zeta\gamma^{-1}$, $M=\gamma T^2\}$, respectively, to get the following non-dimensionalized form:
	\begin{align}
		&\nabla P = \nabla^2\textbf{u}, \label{equ:stokes}\\
		&\nabla P = -\textbf{u}, \label{equ:darcy}\\
		&P_{ext}-P_{int} = -\kappa R,
	\end{align}
	which we then solve.

	To account for behaviors that interpolate between the two extremes with dominating viscosity or dominating friction, the more general equation is obtained by coarse-graining the coupled Langevin equations using the Mori-Zwanzig technique \cite{hessGeneralizedHydrodynamicsSystems1983,evansStatisticalMechanicsNonequilibrium2007}. Considering only the long-wavelength limit and neglecting the hydrodynamic effect of the solvent, the relevant continuum description is
	\begin{equation}
		\rho\frac{D\textbf{u}}{Dt}=-\nabla P+\mu \nabla^2\textbf{u}-\zeta\textbf{u}.
		\label{equ:gNS}
	\end{equation}  
	Here, the $\mu\nabla^2\textbf{u}$ term represents the viscosity originating from the direct interactions between the colloidal particles or biological cells, and $-\zeta\textbf{u}$ represents the additional dissipative effects from the environment (e.g., the frequent collisions of the solvent molecules on colloidal particles, or the frictional effects on cellular systems from substrates or the extracellular matrix). Note that in colloidal systems, the frictional coefficient $\zeta$ is related to the viscosity of the solvent, which should be distinguished from the viscosity $\mu$ we use here.  
	
	We non-dimensionalize this equation into the following form:
	\begin{equation}\label{eq:nonDimGNS}
		\begin{aligned}
			&La\frac{D\textbf{u}}{Dt}= -\nabla P+\nabla^2\textbf{u} -\frac{1}{Da}\textbf{u},\  \textrm{ where}\\
			&La=\rho\frac{R\gamma}{\mu^2}\textrm{,\ \  } Da=\frac{\mu}{\zeta R^2} 
		\end{aligned}
	\end{equation}
	in which the Laplace number ($La$) is the natural coefficient of the inertial term and the inverse Darcy number ($Da^{-1}$) is the natural coefficient of the frictional term (drawing an analogy between $\zeta$ in our case and the permeability $K$ in the usual formulation of Darcy's Law via $\zeta=\frac{\mu}{K}$). Note that $La=Oh^{-2}$, where $Oh$ is the Ohnesorge number typically used to quantify coalescence dynamics \cite{paulsenInexorableResistanceInertia2012a}.

	\section{Microscopic Dynamics Determines Overall Shape Evolution}
	\label{sec:shape}
	
	In Fig.~\ref{fig:Contours}(a-d) we show both representative snapshots of our particle-based simulations and ensemble-averaged contours representing the boundary of the coalescing droplets. In both cases, the total area of the system is conserved. Immediately apparent is how, in the case of MD, as the ``neck'' grows the overall length of the coalescing droplets decreases. This evolution of shapes is in close qualitative agreement with the contour evolution predicted by directly solving the Stokes equation (Eq.~\ref{equ:stokes}) as shown in Fig.~\ref{fig:Contours}(e), which was done for this geometry in the classic work of Hopper \cite{hopperCoalescenceTwoViscous1993,hopperCoalescenceTwoViscous1993a}. In contrast, in our BD simulations we see that the overall droplet length changes very little even while the neck grows substantially. In Fig.~\ref{fig:Contours}(f) we show that this overall evolution of the coalescing droplets appears similar to the result of solving the equation of Darcy's law (Eq.~\ref{equ:darcy}). We also note that a suggestively similar qualitative trend was recently observed in nucleoli fusion \cite{caragineSurfaceFluctuationsCoalescence2018a}. 
	\begin{figure}[ht]
		\centering
		\includegraphics[width=\columnwidth]{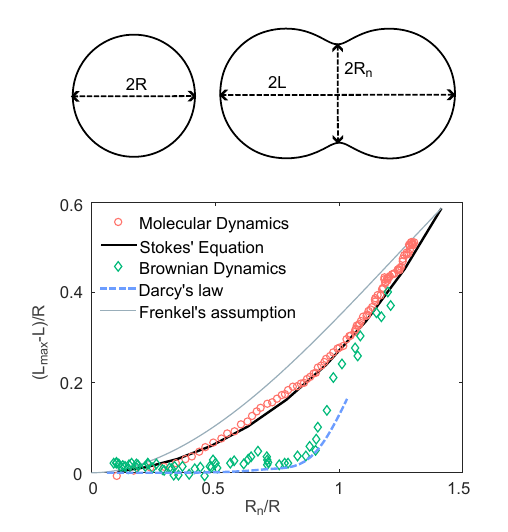}
		\caption{\textbf{Shape evolution curves for different microscopic dynamics.} Upper Panel: Schematic diagram of quantities measured during coalescence: the initial radius of each disk ($R$), the overall length of the coalescing system ($2L$), and the neck radius $R_n$. Lower Panel: Comparison of shape evolution curves predicted by different theoretical approaches and those observed in our MD and BD simulations (error bars omitted). The results for particle-based simulations are of systems with $N=50000$ particles, averaged over 5 (MD) or 3 (BD) independent simulations.} 
		\label{fig:shape evolution}
	\end{figure}
	
	\begin{figure*}[ht]
		\centering
		\includegraphics[width=\textwidth]{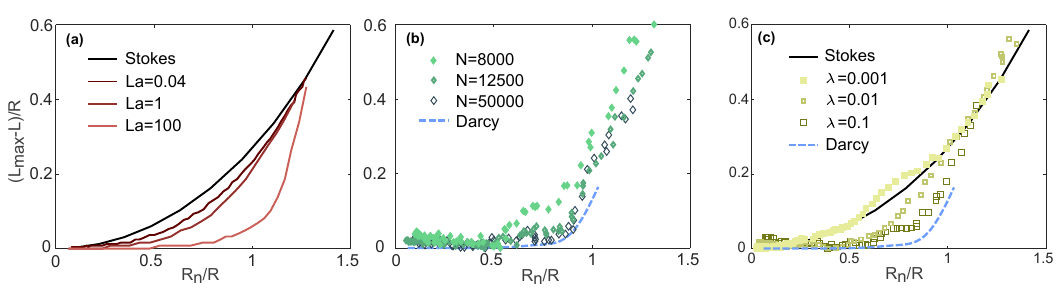}
		\caption{\textbf{Shape evolution curves for droplet coalescence with different parameters in continuum equations or particle-based simulations.} (a) Shape evolution curves for the numerical solutions of the standard incompressible Navier-Stokes equation with different Laplace numbers, compared with Hopper's solution. (b) Shape evolution curves (averaged over 3 simulations for $N=50000$ and 5 simulations for both $N=12500$ and $N=8000$) for the Brownian Dynamics simulations with different system sizes, compared with the solution of Darcy's law. (c) Shape evolution curves (averaged over 5 simulations for each curve) for $N=50000$ simulations using Langevin Dynamics simulations with different values of the damping coefficient $\lambda$.}
		\label{fig:diagram}
	\end{figure*}
	To quantify the differences in the dynamics of droplet coalescence we focus on two geometrical measures: the total length $2L$ and the neck radius $R_n$, as schematically defined in Fig.~\ref{fig:shape evolution}; we measure these in units of the radius of the original drops, $R$. We first focus on the ``shape evolution curve'' of the coalescing droplets in Fig.~\ref{fig:shape evolution}. This is a time-independent parameterization of the overall coalescence process, which we define as the decrease in combined droplet length (quantifying the approaching of the two droplets' centers) as a function of the neck radius. In this representation, our MD simulations are in quantitative agreement with Hopper's solution of Stokes equation. For comparison, we also show the shape evolution curve predicted by an earlier, approximate theory due to Frenkel \cite{frenkel1945viscous}; this theory makes strong assumptions about the coalescing geometry, and is frequently used to compare with experimental data on complex systems \cite{flennerKineticMonteCarlo2012,grosserCellNucleusShape2021a,koshelevaCellSpheroidFusion2020a}. Again contrasting with this, the BD simulations and the solution of Darcy's law show strong deviations from Hopper's prediction, and are characterized by an initial plateau in the shape evolution curve that suggests a delay in the droplets' approach.

	As we mentioned in Sec.~\ref{sec:Methods}, Darcy's law is the overdamped limit of Eq.~\ref{equ:gNS} when the inertia and viscosity are neglected. However, real systems have neither completely vanishing inertia nor completely vanishing viscosity. So, we need to analyze the full equation (Eq.~\ref{eq:nonDimGNS}) to show the effect of competition between inertia, viscosity, and friction. When $Da^{-1}=0$, the full equation reduces to the standard Navier-Stokes equation for simple liquids, and the Stokes' equation is the extreme case when $La$ is also zero. We show the effect of the Laplace number $La$ by solving the standard Navier-Stokes equation in our problem geometry with different values of $\mu$ while keeping $\rho$, $\gamma$ and $R$ constant, as shown in Fig.~\ref{fig:diagram}(a). Increasing the value of $La$ by decreasing the viscosity $\mu$ leads to the elongation of the plateau. Similarly, we can expect from the dependence  of $La$ shown in Eq.~\ref{eq:nonDimGNS}, that increasing the system size $R$ has a similar effect as decreasing $\mu$. This result is supported by the recent MD simulations of Ref.~\cite{heinenDropletCoalescenceMolecular2022a}. In the case of BD when the inertia is totally neglected ($La=0$), we see that a larger value of the Darcy number (which corresponds to a larger role for the viscosity in Eq.~\ref{eq:nonDimGNS}) should lead to results that increasingly \emph{deviate} from the extreme case of Darcy's Law (the case when $Da=0$). This explains the system-size effect seen in the shape evolution curves of Fig.~\ref{fig:diagram}(b). Based on the above results, we see that by tuning parameters in MD (no friction) and BD (no inertia) simulations, we can span the whole range for simple liquid and frictional liquid separately, but cannot transit from one to the other. This can only be achieved with the Langevin Dynamics with coexisting inertia and friction terms and as shown in Fig.~\ref{fig:diagram}(c), by tuning the relative strength between friction and inertia.  
	
	Doing so, we see that increasing either the inertial or frictional effect relative to the viscous effect results in an elongated plateau in the shape evolution curve. This corresponds to increasing the value of $La$ or $Da^{-1}$ in the continuum equation, to increasing the mass $m$, system size, and friction coefficient $\lambda$, or decreasing the inter-particle interactions in discrete simulations. A natural question to ask is: are the explanations for deviations in shape evolution curves (manifested as the length of the plateau) in these quite physically different systems (inviscid or frictional) related? We propose that a simple way to think about this feature of droplet coalescence is in terms of the propagation of perturbations.The plateau in the shape evolution curve implies some lag between the shape change in the neck region and at the two sides (far away from the neck region). The neck starts to grow immediately as the two droplets contact, driven by the greatly changed stress at the contacting point, but it takes time for the information of these changes to the stresses to propagate away from the neck region. One can qualitatively compare the timescale for different modes of propagation, and thus estimate the length of the plateau, via the analysis of the linear Boltzmann equation in kinetic theory. 
	
	\begin{figure}[ht]
		\centering
		\includegraphics[width=\columnwidth]{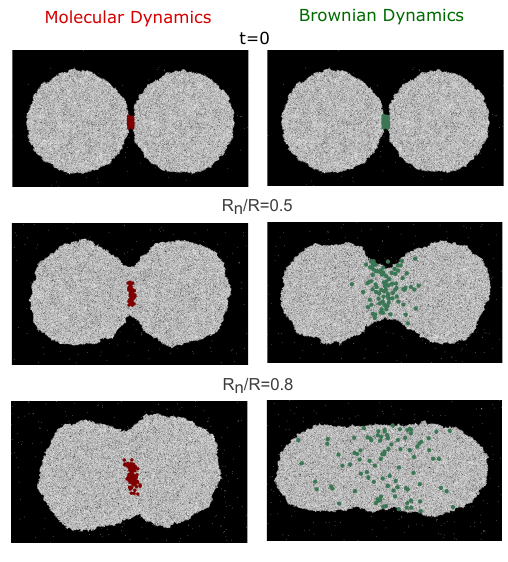}
		\caption{\textbf{Comparison of time scales for shape change and particle diffusion.} At $t=0$, the particles near the neck region are colored (as red in MD simulations and green in BD simulations) and snapshots are taken when $R_n/R\approx 0.5$ and $R_n/R\approx 0.8$.}
		\label{fig:diffuse}
	\end{figure}
	
	For a $d$-dimensional simple liquid, the perturbation analysis of the linear Boltzmann equation gives $d+2$ modes, including $2$ sound modes (wave-like), $d-1$ shear or viscosity modes (diffusive) and $1$ heat mode (diffusive) \cite{soto2016kinetic,dorfman2021contemporary}. Sound modes are reversible and the other modes account for the irreversible transport of conserved quantities (momentum for shear/viscosity modes and mass for heat mode) in the process towards equilibrium. The diffusion coefficients for the shear modes are the kinematic viscosity $\nu = \frac{\mu}{\rho}$ while the heat mode corresponds to the thermal diffusion of particles. For a frictional liquid, momentum is no longer conserved and only the heat mode remains \cite{hessGeneralizedHydrodynamicsSystems1983}. In Fig.~\ref{fig:diffuse} we visually illustrate the difference the presence or absence of these modes makes. We color the particles near the neck region at the beginning of the simulations with red or green for MD or BD respectively, and then take snapshots at the time points when the length begins decreasing, with normalized neck radii $R_n/R$ reaching $0.5$ and $0.8$ in the two cases. We see that for Brownian Dynamics, the time scale needed to initiate a decrease in length corresponds to that needed for the particles to diffuse to the two sides. In contrast, for Molecular Dynamics (high viscosity case), as the kinematic viscosity is large, the momentum transport process is much faster than the particle diffusion process. For the Stokes equation with $\rho$ being neglected, the kinematic viscosity is infinite, resulting in the extreme case with no plateau in Fig.~\ref{fig:shape evolution}(a), and the decrease in length and increase of neck radius start simultaneously. With a larger Laplace number (smaller kinematic viscosity), the momentum transport through the shear mode becomes slower, resulting in a longer plateau shown in Fig.~\ref{fig:shape evolution}(a). This effect of decreasing viscosity is also observed in experiments \cite{paulsenInexorableResistanceInertia2012a} and Molecular Dynamics simulations in \cite{heinenDropletCoalescenceMolecular2022a}.  
	
	\section{Novel Scaling Laws of Neck Growth}
	\label{sec:scale}
	Another simple yet powerful way to quantify the coalescence of two droplets is through the time evolution of the neck radius. Simple liquids have universal exponents characterizing this time evolution in both the initial and late stages of coalescence. This universality can be understood via the analytical study of the Stokes equation by Hopper \cite{hopperCoalescenceTwoViscous1993, eggersCoalescenceLiquidDrops1999a}, by direct numerical simulations of the Navier-Stokes equation \cite{menchaca2001coalescence,anthony2020coalescence}, and by qualitative theoretical approximations that use force balance at the droplet interface \cite{eggersCoalescenceLiquidDrops1999a} -- all of which are well-supported by experimental data \cite{wu2004scaling, paulsenViscousInertialCrossover2011, paulsenInexorableResistanceInertia2012a,paulsen2013pre, paulsenCoalescenceBubblesDrops2014} and MD simulations  \cite{perumantath2019MD, pothier2012molecular}. 
	\begin{figure}[ht]
		\centering
		\includegraphics[width=\columnwidth]{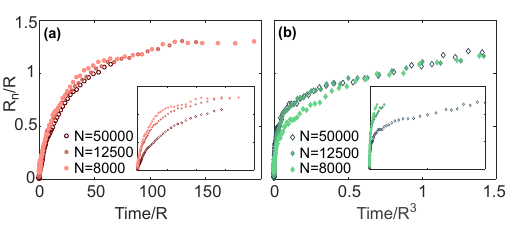}
		\caption{\textbf{The dependence of time scale on system size.} (a) Neck growth curves for Molecular Dynamics simulations with different system sizes. The time is rescaled by the initial radius $R$ for one droplet. (b) Neck growth curves for Brownian Dynamics simulations with different system sizes. The time is rescaled by the cube of the initial radius $R$ for one droplet. The insets are the curves before rescaling the time.}
		\label{fig:systemsize}
	\end{figure}
	
\begin{figure}[ht]
		\centering
		\includegraphics[width=\columnwidth]{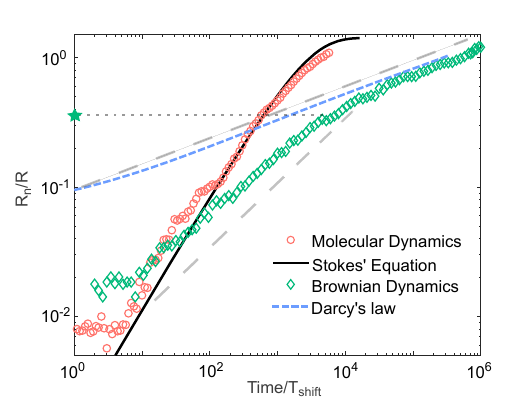}
		\caption{\textbf{Neck growth scaling for different microscopic dynamics and continuum equations.} Log-log plot of the time evolution of the neck growth. The dashed gray lines are guides to the eye with slopes $1/2$ and $1/5$. The time axis has been rescaled for different dynamical systems to allow for convenient comparison, and the green star is an estimated crossover point for the BD data. The results for particle-based simulations are of systems with $N=50000$ particles, averaged over 5 (MD) or 3 (BD) independent simulations.}
		\label{fig:scaling}
	\end{figure}

	One simple aspect of the overall time dependence is that the characteristic time of coalescence for viscous liquids, $\tau_{viscous}\propto R$, can be inferred from the rescaling to obtain Eq.~\ref{equ:stokes}. However, for a frictional liquid, based on the rescaling to obtain Eq.~\ref{equ:darcy}, the characteristic time in $\tau_{frction}\propto R^3$. So, the coalescence of frictional liquid has a much stronger dependence on system size, as shown in Fig.~\ref{fig:systemsize}. A more important feature is the universal scaling laws governing neck growth \cite{xia2019universality}. In simple liquid systems the scaling of the neck growth is determined by a viscous regime at early times, in which $R^{viscous}_n \sim t$ or $R^{viscous}_n \sim -t\log t$, depending on the viscosity and geometrical details of the singular contact point  \cite{paulsenInexorableResistanceInertia2012a,anthony2020coalescence}. These dynamics can give way to an inertia-dominated regime at late times, in which $R_n^{inertial} \sim t^{1/2}$. In our MD simulations of highly viscous drops, we do not observe the late-time crossover to an inertial regime (Fig.~\ref{fig:scaling}), and instead the dynamics closely follow the well-known, purely viscous, coalescence solution of Stokes' equation.
	
	In contrast, our BD simulations show a much slower growth of the neck radius with time. They also appear to have two dynamical regimes, with a scaling law (over a limited range) close to $t^{1/2}$ at \emph{early} times crossing over to a scaling law whose exponent is obviously \emph{smaller} than $t^{1/2}$ (approximately $t^{1/5}$) at later times. We first point out that the late regime with $t^{1/5}$ scaling law is consistent with the direct prediction of Darcy's law (obtained by our boundary integral analysis). This is shown by the blue dashed lines in Fig.~\ref{fig:scaling}, and is consistent with the theoretical picture we describe above. The early regime does not have a clear analog with any of the known regimes for simple liquids governed by the Navier-Stokes equation or frictional liquids obeying Darcy's law. 
	
	\begin{figure}[h]
		\centering
		\includegraphics[width=\columnwidth]{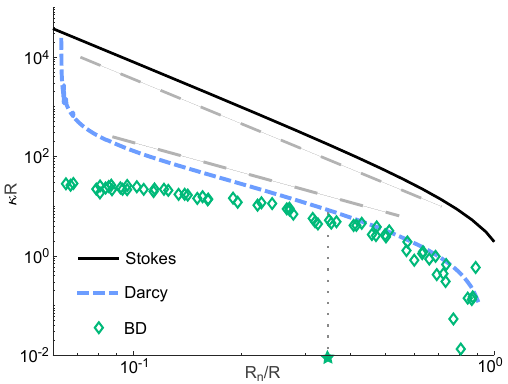}
		\caption{\textbf{Curvature limit in Brownian Dynamics due to finite particle size causes the transition between two regimes in neck growth.} Log-log plot of the curvature of the droplet boundary at the neck as a function of neck radius. The dashed gray lines are guides to the eye with slope $-2$ and $-3$, and the green star is the estimated crossover point at which the BD's curve of curvature starts to deviate from that for Darcy's law. The results for particle-based simulations are of systems with $N=50000$ particles, averaged over 3 independent simulations.}
		\label{fig:curvature}
	\end{figure}
	
	To understand this deviation of BD simulations from the continuum theory (Darcy's law), we investigate the local curvature $\kappa$ at the neck since the stress caused by surface tension, set by $\gamma\kappa$, is the driving force behind coalescence. In Fig.~\ref{fig:curvature} we plot the  rescaled maximum curvature $\kappa R$ (located at the neck) versus $R_n/R$. We find, with no adjustable parameters, that the curvatures from Darcy's Law and our BD simulations agree in the late regime and both are more than an order of magnitude smaller than the curvature for Stokes coalescence. This explains the slow growth ($t^{1/5}$ regime) in Fig.~\ref{fig:scaling}. Similar to the neck growth curves, a deviation between BD simulations and Darcy's law also occurs at early stage when the curvature is strongly affected by the initial conditions. The abrupt drop of curvature from very high value at the very beginning of the coalescence in the solution of Darcy's law is caused by the artificial choice of initial geometric shape -- we use the analytical solution of Stokes equation at $R_n\approx0.06$, since the analytic form of the solution to Darcy's law at early times is not known for this geometry. 
 
 Note that the relatively low curvature at the early stage in our BD simulations is physically meaningful. In particulate simulations, the curvature at the neck is limited by the finite size $r_e$ of the particles themselves (i.e., the normalized curvature $\kappa R$ in our particulate simulations is bounded by $\sim R/r_e$). The lower curvatures than continuum solutions at the early stage help rationalize the existence of the unexplained $t^{1/2}$ regime we find in Fig.~\ref{fig:scaling} (although not the precise exponent). As a validation of this explanation, we mark the crossover points (with green stars) between the two regimes in both Fig.~\ref{fig:scaling} and \ref{fig:curvature} and they are roughly at the same position ($R_n/R\approx0.3\sim 0.4$). This explanation also predicts that systems with a smaller ratio between system size and single element size $R/r_e$ will have a greater finite size effect and thus a later crossover point, as shown in Fig.~\ref{fig:crosspoint} in the Appendix. Reading from Fig.~\ref{fig:curvature}, when  $R/r_e\geq 10^3$, the crossover point will be smaller than $R_n/R\approx 0.1$ and thus hard to detect. For molecular liquids, the characteristic particle size is $\sim 0.1$ nm so that $R/r_e\sim10^{7}$ when the droplet size is $\sim$ mm, which can be confidently regarded as continuum. However, many complex liquid systems have much larger relative particle sizes. For example, a cellular spheroid usually consists of less than $10^4$ cells, which gives $R/r_e\sim 20$; the size of a nucleolus is $\sim 2~\mu $m \cite{caragineSurfaceFluctuationsCoalescence2018a} and is mainly composed of large molecules (using RNA polymerase as a representative with size $\sim 10$ nm \cite{milo2015cell}) so that $R/r_e\sim 200$. In these systems, local behaviors within a small region, such as the early phase neck region in the coalescence case, might not be explained with purely continuum theory. 
	
We dedicate substantial effort to the explanation of the new $t^{1/2}$ regime in coalescing frictional liquids because the scaling law is similar to the inertia-dominated regime in the late stages of simple liquid coalescence, and thus can be misconstrued. For instance, when a scaling law close to $\sim1/2$ is observed within some range in an experiment, one could estimate the inside/outside viscosity ratio assuming the dynamics of a simple liquid \cite{paulsenCoalescenceBubblesDrops2014}. Instead, our work shows that there is another regime of with a scaling law $\sim1/2$ whose origin is distinct from the same scaling regime in simple liquids. We return to this point with a concrete example in in Sec.~\ref{sec:significance}.      
	
	\section{Summary and Future Directions}  
	\label{sec:future}
 \begin{figure}[h]
		\centering
		\includegraphics[width=\columnwidth]{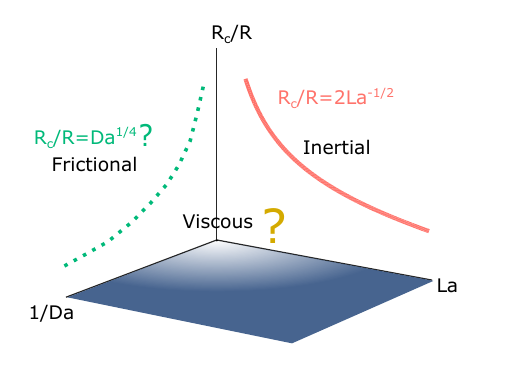}
		\caption{\textbf{Proposed phase diagram showing the transition between different coalescence regimes}. The red solid line shows the transition between viscous (below) and inertial (above) regimes in the simple liquid case and the green dotted line shows \emph{qualitatively} the transition between viscous (below) and frictional (above) regimes for a frictional liquid. The region close to the origin is dominated by viscosity. The color gradient from light blue to dark blue represents the qualitative increase in the plateau length from the shape evolution curves.}
		\label{fig:summary}
	\end{figure}
	In this work, we combined particle-based simulations and continuum theories to investigate the coalescence of liquid droplets according to different microscopic dynamics. Except for the finite element size effect, the other contents can be summarized in Fig.~\ref{fig:summary} and Table~\ref{tab:summary}. On the $1/Da-La$ plane, the gradient of color from light to dark qualitatively represents the trend of increasing plateau in the shape evolution curve as shown in Fig.~\ref{fig:shape evolution} and \ref{fig:diagram}. For the scaling laws of neck growth, what was thoroughly studied previously is the transition from viscous regime to inertial regime on the $R_c/R-La$ plane. Through a pioneering series of studies \cite{eggersCoalescenceLiquidDrops1999a,paulsenViscousInertialCrossover2011,xia2019universality}, a quantitative description of crossover neck radius between the two regimes was obtained as $R_c/R = 2/\sqrt{La}$. One can obtain this relation by setting $La = \frac{ L\gamma}{\mu^2}=1$ with the correct choice of lengthscale $L=R_c^2/R$, which comes from the largest curvature at the minimum neck radius. 
 Prior to this, a quantitatively incorrect but qualitatively reasonable choice, $L=R_c$, was used. For the frictional case ($La=0$), the local geometry and flows near the neck region are still unclear and thus we cannot predict accurately the crossover neck radius. But a similar procedure can be implemented to get a rough trend, $R_c/R \propto Da^{1/4}$, by setting $Da=\frac{\mu}{\zeta L^2}=1$ and choosing $L=R_c^2/R$ again. This is plotted on the $R_c/R-1/Da$ plane in Fig.~\ref{fig:summary}. In this work we have focused most of our effort on the purely frictional case, and the comparison to the other two limiting cases (purely viscous and inertial) is listed in Table~\ref{tab:summary}. However, the intermediate cases with all three terms (inertia, viscosity, and friction) remains insufficiently explored due to the difficulty in solving the full equation (\ref{eq:nonDimGNS}). This will be an important future direction as many physical and biological systems lie in the intermediate regime.   

 For the frictional case, we demonstrated several experimentally-testable qualitative and quantitative features that can distinguish the overall shape evolution of the system with overdamped dynamics from the theoretical predictions derived in the context of Stokes flow, and further showed that we expect unusual power laws characterizing the neck growth. Several natural extensions of this computational and theoretical work suggest themselves. Most immediately, tests of our results can further be obtained by connecting Darcy's Law to semi-2D flows in Hele-Shaw cells (confined in between two plates) \cite{mclean_saffman_1981}. Some existing experimental results in this context seem consistent with our prediction for the overall shape evolution (e.g., Ref.~\cite{brun2013generic}). In such an experiment we would expect neck growth scaling laws of $R_n\sim -t\log t$ at early stages (when the neck radius is smaller than or comparable to the gap width of the Hele-Shaw cell); as the neck grows and the system becomes better described by Darcy's Law rather than the Stokes equation, we would expect a new scaling of $R_n\sim t^{\alpha}$. Thus, distinguishing the frictional coalescence regime we propose requires a small enough gap size between the two plates. These two regimes may have been observed in the experiments of Ref.~\cite{yokota_dimensional_2011}, although the interpretation is complicated by their use of a vertical Hele-Shaw cell to study a drop coalescing into a planar interface. 
\begin{table}
     \centering
     \begin{tabular}{c|c|c|c}
          & $\alpha$ & $\tau$ & droplets approaching\\
         \hline
          Frictional &  $\sim1/5$ & $\sim R^3$ & Delayed\\
          Viscous &  $1$ & $\sim R$ & No Delay\\
          Inertial & $1/2$ & $\sim R^{3/2}$ \cite{heinenDropletCoalescenceMolecular2022a} & Delayed \\
     \end{tabular}
     \caption{\textbf{Comparison of coalescing dynamics between purely frictional, viscous, and inertial cases.} $\alpha$ is the scaling law for neck growth, and $\tau$ is the characteristic time. Since the pure inertial case is less relevant for the low Reynold number systems we are interested in and was studied in \cite{heinenDropletCoalescenceMolecular2022a}, we simply provide the results here without elaborating. }
     \label{tab:summary}
 \end{table} 	
	We have mentioned the main assumptions underlying our discrete simulations in Sec.~\ref{sec:Methods}, which could potentially impose some limitations on our findings. In particular, true hydrodynamic interactions are long-ranged ($1/r$ or $1/\log{r}$) while the generic Lennard-Jones potential we use is short ranged. This might not have a large effect in the bulk after hydrodynamic screening is accounted for, but may have a bigger influence on the neck growth since particles lie near the interface. As we are not focusing on fast dynamics at the early stage, the qualitatively different behaviors we find in the coalescence of frictional liquids should still hold, but we will investigate the effect of long-range interactions at the surface in future work. We additionally comment that we do not know a priori whether the results for 3D systems are the same as our 2D results: the early stage of the coalescence of simple liquids is not affected by dimensionality, but this has not yet been shown to hold for the frictional coalescence processes considered here. We find strong, qualitative support for our findings in the overall shape evolution from 3D simulation results with BD in \cite{flennerKineticMonteCarlo2012} (which considered direct BD and Kinetic Monte Carlo simulations, both of which produce overdamped dynamics). This suggests that the new behaviors we find are due to the general feature of overdamping, rather than details of simulation methods. Studying the further changes in coalescence caused by system-specific features, such as the anomalous mechanical properties and dynamics caused by non-metric cell-cell interactions \cite{sussman2018soft, sussman2018anomalous}, are another natural avenue to pursue.
	
	\section{Experimental Consequences of Frictional Fluid Mechanics} 
	\label{sec:significance}
 
	Beyond these directions for future research to probe the theory of frictional coalescence, we wish to highlight several examples from across a range of biological and soft matter experiments where we believe the effects we predict will help uncover the basic physics of their respective systems. These examples include structures within cell nuclei composed of macromolecules, and multicellular aggregates such as tumor spheroids and bacteria colonies. In each of these systems, coalescence is a critical process or has been used as a tool to infer important physical properties or biological mechanisms. We have shown above that simple liquid theories or models may not be applicable in these systems. Our work provides new interpretations for some of the previously reported experimental results and should be considered in future studies to enhance our understanding of these systems. Thus, we suggest that the implications of our work extend beyond the study of coalescence and can be applied to other processes in ``dry'' soft matter systems.
	\begin{figure}[h]
		\centering
		\includegraphics[width=\columnwidth]{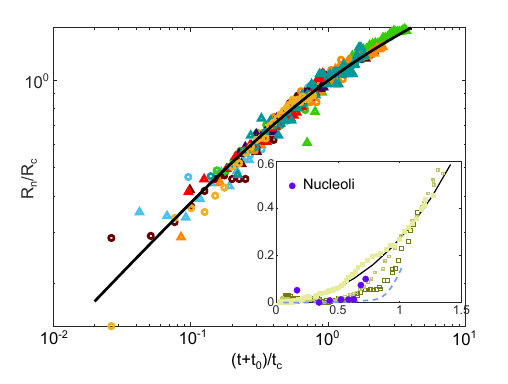}
		\caption{\textbf{Experimental results on the coalescence of nucleoli (Adpated from Fig.~1d and Fig.~4a of Ref.~\cite{caragineSurfaceFluctuationsCoalescence2018a}).} Neck growth scaling from 14 fusing events are shown with different markers and the black solid line is the fitted curve $y = 2(x^{-0.5}+x^{-0.2})^{-1}$ with $x=(t+t_0)/t_c$ and $y=R_n/R_c$. [Inset] Shape evolution curves extracted from contours presented in Fig.~1(d) of Ref.~\cite{caragineSurfaceFluctuationsCoalescence2018a} are compared with Langevin Dynamics simulation results from Fig.~\ref{fig:shape evolution}c above.}
		\label{fig:nucleoli}
	\end{figure}
	
	We first discuss an example from sub-cellular systems. The mechanical properties of a cell's nucleus and its microstructures play a critical role in many cellular processes \cite{isermann2013nuclear,friedl2011nuclear}. Therefore, measuring these properties is of great importance. A non-invasive approach to measuring physical properties of the nucleoli and the nucleoplasm, such as their viscosity, was proposed by measuring the natural coalescence process of the nucleoli \cite{caragineSurfaceFluctuationsCoalescence2018a}. This work assumed that the nucleoli behave like viscous liquid droplets and inferred a high external/internal viscosity ratio based on the scaling law of neck radius growth over time. Examining the overall shape evolution contours in Fig.~1 of that works suggests a process more akin to that of a frictional fluid, with a pronounced (although not as long as the extreme case of Darcy's law), time-delayed decrease of length compared to growth of the neck, as shown in the inset of Fig.~\ref{fig:nucleoli}. 
 
 The delay in length decrease suggests that the inverse Darcy number for this system, quantifying the relative competition between friction and viscosity, is non-negligible but also not large enough to dominate. This is reasonble and consistent with previous work, as nucleoli are often modeled as a polydisperse colloidal system with macromolecules. More data on the contours during fusion -- or other direct experimental measurements -- would be required to accurately estimate the Darcy number here, but this already suggests that the interpretation of the observed scaling law as stemming from a high external/internal viscosity ratio may not be the proper explanation. Our results predict, in addition to this $1/2$ scaling law, a late-time transition to a new scaling law with an exponent close to $1/5$. Hints of this are visible in Fig.~4b of Ref.~\cite{caragineSurfaceFluctuationsCoalescence2018a}, but since the coalescence of frictional liquids was at that time unexplored the authors did not focus on this regime. We note that the normalized neck radius measured in their experiments is always lower than $1$ so that the coalescence has not reached the stage slowed down by geometric restrictions ($R_{\text{final}}/R=\sqrt{2}$) and thus we believe that the smaller scaling law at the late stage in that figure is a nontrivial indication of relevant physics.
 
 To highlight this, we extracted data from their Fig.~4a and fit them with 
 \begin{equation}
     R_n  = 2R_c\left[\left(\frac{t+t_0}{t_c}\right)^{-0.5}+\left(\frac{t+t_0}{t_c}\right)^{-0.2}\right]^{-1},
 \end{equation}
a functional form which interpolates between our two predicted scaling laws  of $1/2$ and $1/5$). Here $t_0$, $t_c$ and $R_c$ are fit parameters, with $t_0$ restricted in the range the same as in the original paper (i.e., between $20 - 310s$). We see in Fig.~\ref{fig:nucleoli} that all the data collapse onto a single curve. Additionally, the fitted crossover neck radius $R_c$ is on average $\sim 0.48R$, which is reasonable considering that the rough size ratio between the nucleolus and RNA polymerase ($\sim 100$) is close to our BD simulations with 50000 particles. This crossover point – set by $R/r_e$ – might imply a natural bound on droplet size distributions for these mesoscopic nucleolar droplets which naturally grow by colliding and coalescing, as droplets larger than the crossover scale would coalesce in a much slower manner. The above discussions are only qualitative or semi-quantitative due to the following factors: (1) only one set of contours was presented as a visual example; (2) most of the neck growth measurements were stopped before or near the crossover point between two regimes; and (3) $t_0$, which greatly influences the scaling law, is fitted instead of measured. Nevertheless, our work facilitates a natural explanation for the observed scaling behavior, and provides valuable information on potential future experimental research.
	
	\begin{figure}[h]
		\centering
		\includegraphics[width=\columnwidth]{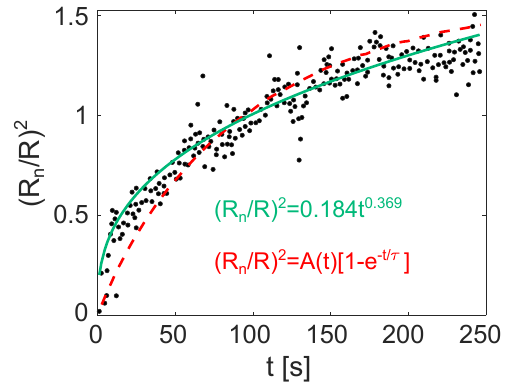}
		\caption{\textbf{Experimental results on the coalescence of bacteria colonies (Adpated from Fig.~S10b of Ref.~\cite{welker2018molecular}).} The red dashed line is their fitting with simple liquid theory and the green solid line is our fitting with the power law function $(R_n/R)^2=At^\alpha$, where $\alpha=0.369$, consistent with the scaling law $1/5$ for $R_n/R$ in our frictional liquid theory. }
		\label{fig:bacteria}
	\end{figure}
 
	We next discuss examples at a much larger length scales: bacterial colonies and cellular aggregates. The coalescing dynamics in these systems provide valuable insights into the macroscopic mechanical properties and microscopic cell behaviors, as well as their relations, within these systems \cite{welker2018molecular,ponisch2018pili,grosserCellNucleusShape2021a, koshelevaCellSpheroidFusion2020a}. Such studies enable the development of methods to effectively tune and control these complex systems as a whole, employing well-developed genetic techniques. For example, attraction between bacteria can be controlled by genetically manipulating pilus motor activity, leading to an effective surface tension and viscosity that governs an entire bacterial colony \cite{welker2018molecular}. These physical parameters have been inferred by comparing fusing colonies with the predictions of simple liquid theory, but a discrepancy for the short-term fusion dynamics can be readily seen in Fig.~\ref{fig:bacteria} between dots and the red line (reproduced using data from Fig.~S10b in \cite{welker2018molecular}).
 
 The authors of \cite{welker2018molecular} also noticed this discrepancy and proposed an explanation involving a special motile layer of bacteria at the colony surface, which in turn implicates new sensing and motility-regulating pathways. We find, however, that the early-time fusion dynamics are readily explained by our theory of frictional liquid coalescence (green line in Fig.~\ref{fig:bacteria}) without the need of introducing specialized bacteria. A power law consistent with our predictions, $R_n/R\propto t^{0.185}$, can fit the neck growth curves without the systematic deviation in the early stage. So, the fusion dynamics they observed could be natural consequence of the frictional effect between bacteria and the environment, and the introduction of specialized cells may not be necessary. As inertia is typically negligible compared to friction for moving bacteria, a frictional liquid theory (at the continuum level) or models grounded in Brownian Dynamics (at the bacterial degree-of-freedom level) is a more suitable theoretical framework. Moreover, it is known that surface friction determined by the stiffness of substrates can influence the morphology and growth of bacteria colonies \cite{fei2020nonuniform, asp2022spreading}. So, introducing a parameter governing the frictional effect as in our frictional liquid theory offers a framework for studying these effects.

Experimental observations on spheroids composed of cells have revealed distinct and distinctive fusion dynamics in healthy versus cancerous cell types~\cite{grosserCellNucleusShape2021a, koshelevaCellSpheroidFusion2020a}, with qualitative connections drawn to microscopic characteristics like cell shape and extracellular matrix composition. Although these qualitative comparisons have offered valuable insights into the physical differences between cell types, the absence of adequate theoretical frameworks (made more daunting by the clear, system-specific and non-universal properties of cells) has hindered the formulation of testable quantitative predictions and practical guidelines. However, what is clear is that the simple viscous liquid theory, based on Stokes equation or its approximations, is not the correct foundation for many of these systems with more biological complexities. For example, an essential aspect of biological systems lies in their out-of-equilibrium nature. Many of the most common models of simple active matter (meant to describe systems ranging from dictyostelium, mouse fibroblasts, and human endothelial cells \cite{selmeczi2008cell,dunn1987unified,stokes1991migration}) begin with overdamped agents with an active force (e.g., from an Ornstein-Uhlenbeck process) breaking equilibrium. In these models the ratio of ``friction'' and ``inertia'' ratio in phenomenological Langevin-equation descriptions can be related to the persistence time governing the active forces. Our findings on frictional fluids can serve as a limiting case when the cells' active movement is inhibited. Of course, due to complex intra- or inter-cellular signaling pathways controlling the polarization of moving cells \cite{skoge2014cellular, yue2018minimal}, cellular movement cannot always be phenomenologically be described by a simple Langevin Equation \cite{wu2014three, nousi2021single}. But no matter how complicated they are, one common feature is the high dissipation. Therefore, our work, which incorporates the role of friction and fluctuation, serves as a better base model than the simple liquid theory, upon which more intricate models can be constructed to elucidate the more complex phenomena observed in various cellular tissues.

	We emphasize that the findings presented in this paper extend beyond the scope of coalescence phenomena. Although it is well-known that wet and dry hydrodynamics can lead to different pattern formation and modes of self-organization \cite{marchetti2013hydrodynamics}, our study emphasizes that coupling to dissipative environments can alter the fundamental behavior, even the universality class, governing a range of physical processes. This calls attention to the effect of friction in diverse systems. For example, stronger damping in a driven-dissipative particle system induces a transition from simple equilibrium fluids to non-equilibrium hyperuniform fluids \cite{lei2019hydrodynamics}. Remarkably, the hydrodynamic theory for hyperuniform fluids presented in this study resembles our generalized Navier-Stokes Equation with an additional damping term. In addition to the different collective patterns found in ``dry" and ``wet" systems \cite{doostmohammadi2016stabilization}, another important avenue of research in active matter involves the phase separation behavior of active Brownian particle systems and the non-regular interface properties between the phases \cite{bialke2015negative}. However, such studies are in systems far from equilibrium with artificially defined rules for the energy injection that breaks the fluctuation-dissipation theorem. Our work emphasizes that even in an equilibrium system with a heat bath to compensate for the energy dissipation in a way obeying the fluctuation-dissipation theorem, hydrodynamic behaviors like coalescence can be non-regular, and that the effect of driven activity and the effect of dissipation can and should be decoupled. In a separate study, we have discovered that in models of dense cellular systems Brownian Dynamics alone can give rise to non-regular interface fluctuations in the liquid phase \cite{YueInPrep2023}. This again underscores the necessity for novel continuum theories that transcend the behavior of simple liquids. To summarize, our work not only provides a fresh interpretation of experimental observations on coalescence in diverse frictional liquid systems but also paves the way for a broader research direction that emphasizes the role of frictional effects in colloidal and biological systems writ large.

	\begin{acknowledgments}
		This material is based upon work supported by the National Science Foundation under Grant No.~DMR-2143815 (DMS) and the  NSF iPoLS Student Research Network, Grant 1806833 (HY). The particle-based simulations used the Extreme Science and Engineering Discovery Environment (XSEDE) with a startup allocation (PHY210055), which is supported by National Science Foundation grant number ACI-1053575.
	\end{acknowledgments}
	
	\appendix
	\label{sec:appendix}
	\section{Solving the continuum equations}
	\subsection{Navier-Stokes Equation}
	Our numerical solutions of the Navier-Stokes Equation: $\rho\frac{D\textbf{u}}{Dt}=-\nabla P+\mu \nabla^2\textbf{u}$ are obtained using the open-source \emph{Basilisk} software package \cite{basilisk} with the same parameters for inner and outer fluid: $\rho=1$, $\gamma=1$ and $\mu$ as written in the figure in the main text.
	
	\subsection{Darcy's law}
	\begin{figure}[h]
		\centering
		\includegraphics[width=\columnwidth]{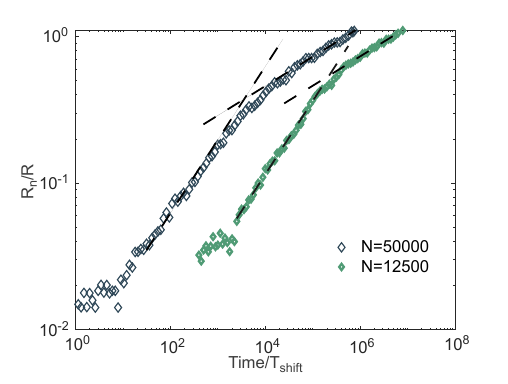}
		\caption{\textbf{Crossover between two regimes in the neck growth curves in discrete simulations with different system sizes.} The dashed lines are the results of linear regression with fixed slope $1/2$ and $1/5$ respectively, on log-transformed data within the proper range. The range is determined by choosing the best fit. We manually shift the curve of $N=12500$ to the right to make it separate from the $N=25000$ curve. }
		\label{fig:crosspoint}
	\end{figure}
	Due to the well-known appearance of spurious vorticities (and other numerical artifacts) in continuum simulations of systems controlled by surface tension  \cite{popinet2018numerical}, we solve Darcy's law, $\nabla P = -\zeta \textbf{u}$, for the coalescing droplets' geometry via the Boundary Integral Method \cite{pozrikidis2011introduction, burtonTwodimensionalInviscidPinchoff2007}. For the case of two discs of frictional fluid 1 with frictional coefficient $\zeta_1$ coalescing in a background frictional fluid 2 with $\zeta_2$, we can get the distribution density of Green's function dipoles on the interface as:
	
	\begin{equation}
		q(\textbf{x}_0)=-\dfrac{2(C-1)}{C+1}I(q(\textbf{x}_0))+\dfrac{2}{C+1}\kappa(\textbf{x}_0),
	\end{equation}
	where $I(q(\textbf{x}_0))=\int\nabla G({\bf x}_0,{\bf x})\cdot\hat{\bf n}({\bf x})q({\bf x})dS$, $C=\zeta_2/\zeta_1$, and $\kappa(\textbf{x}_0)$ is the curvature.

	For $C\neq1$ this leads to an integro-differential equation that typically requires costly iterative methods to solve; this is especially true in the context of the highly singular initial geometry of two barely-contacting disks in our study. Based on previous theories and experiments \cite{eggersCoalescenceLiquidDrops1999a,paulsenCoalescenceBubblesDrops2014}, we expect the fundamental behavior of the system to be similar for a very broad range of inner- and outer-friction coefficients and thus focus on the computationally simpler $C=1$ case.

	We carry out this computation using custom MATLAB 2020b code. We use 1200 points on the interface which we distribute at each update based on the arc-length $ds$ between neighboring marker points. We use a relation such that $ds \propto (x^2+y^2)^k$, to ensure a denser distribution of points near the neck region (where the curvature and, hence, the need for numerical stability, is highest). $k=1$ at the early stage and is relaxed to $1/2$ as the neck grows and curvature decreases. The interface is advanced using an adaptive stepsize fifth-order Dormand-Prince method \cite{press2007numerical}, and for an initial shape we use $r(\theta)=\sqrt{4(a-\sin^2{\theta})}$ with $a=1.001$, which a polar-coordinate representation of Hopper's solution for the Stokes equation \cite{hopperCoalescenceTwoViscous1993} at a very early time.

	\section{Measuring curvatures}
\begin{figure}[ht]
		\centering
		\includegraphics[width=\columnwidth]{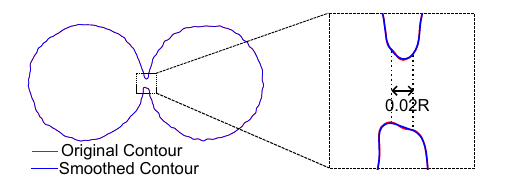}
		\caption{\textbf{Measurement of Curvatures for discrete simulations}An example contour of the Brownian Dynamics simulation (red) and its corresponding smoothed contour (blue) and the range of neck region in which we measure and average the curvature.}
		\label{fig:measurecurvature}
	\end{figure}

	The curvature at the neck is measured based on the average contour of three simulations of Brownian Dynamics with $N=50000$. We first parameterize the contour in terms of arc length $s$, starting from and ending at the left side. Then we use the \textit{csaps} function in MATLAB to smooth the contours with the cubic spline. The smoothing parameter is carefully tuned to make sure that the neck region is not over-smoothed. One example of the smoothing result is shown in Fig.~\ref{fig:measurecurvature}. The curvature is averaged over the neck region with a width equaling $0.02R$ (the radius of particles in this system is about $0.01R$), as shown in Fig.~\ref{fig:measurecurvature}. This range can be varied from $0.01R$ to $0.05R$ without significantly changing the curvature measurement. Curvature at each point is calculated using the formula $\kappa = \frac{x'(s)y''(s)-y'(s)x''(s)}{(x'(s)^2+y'(s)^2)^{3/2}}$. For the numerical solution of Darcy's law, the curvature is measured in a similar way, but without the contour average and only for the minimum neck radius point. 

	\bibliography{brownianCoalescence}
	
\end{document}